\documentclass[a4paper,fleqn]{cas-dc}

\usepackage[T1]{fontenc}
\usepackage[authoryear,round]{natbib}
\usepackage{amsmath,amssymb,amsfonts}
\usepackage{algorithmic}
\usepackage{graphicx}
\usepackage{textcomp}
\usepackage{bmpsize}
\usepackage{xcolor}
\usepackage{booktabs}
\usepackage{multirow}
\usepackage{array}
\usepackage{subcaption}
\usepackage{xurl}
\usepackage{pifont}
\usepackage{tabularx}

\newcommand{\cmark}{\ding{51}}
\newcommand{\xmark}{\ding{55}}

\begin{document}
\let\WriteBookmarks\relax
\def\floatpagepagefraction{1}
\def\textpagefraction{.001}

\shorttitle{SPECTER}
\shortauthors{H. Choi et~al.}

\title[mode=title]{SPECTER: Covert Data Exfiltration from Air-Gapped Systems via USB Power Modulation}

\author[1,2]{Hyeongjun Choi}[orcid=0000-0003-1700-3729]
\ead{tiger8135@korea.ac.kr}
\ead{hjchoi@78researchlab.com}

\author[1]{Ji Hyuk Jung}[orcid=0000-0002-8076-4742]
\ead{graycat@korea.ac.kr}

\author[1]{Ji Won Yoon}[orcid=0000-0003-2123-9849]
\cormark[1]
\ead{jiwon_yoon@korea.ac.kr}

\affiliation[1]{
    organization={Korea University},
    city={Seoul},
    country={South Korea}
}

\affiliation[2]{
    organization={78ResearchLab},
    city={Seoul},
    country={South Korea}
}

\cortext[cor1]{Corresponding author}

\begin{abstract}
We present SPECTER, a post-compromise covert channel that uses host-side USB power control to exfiltrate information from air-gapped and network-segregated systems. A transmitter with administrative privileges encodes bits through USB power-management or device-state operations without placing the payload in USB data transfers. The resulting signals are recovered through three alternative receiver paths based on smartphone charging states, visible peripheral behavior, and near-field RF activity. Experiments on Windows~10 and Ubuntu~22.04 demonstrate the channel on commodity hardware, with charging-state signaling at 0.2--0.25~bit/s across Android and iOS devices and peripheral-state signaling at 1.0~bit/s using a USB mouse on Windows. RF reception is demonstrated using a software-defined radio with a loop antenna in direct contact with the USB cable. Channel availability and observable side effects depend on the host configuration, connected device, and receiver path. These results identify host-side USB power control as a mechanism for low-rate exfiltration of short secrets and motivate treating USB power-management and device-state events as security-relevant telemetry.

\end{abstract}



\maketitle

\section{Introduction}
Air-gapped and network-segregated systems are deployed to reduce the risk of unauthorized data transfer even when conventional defenses such as firewalls, intrusion detection systems, and encryption are in place~\citep{bellovin1994network,khraisat2019survey,popek1979encryption}. Once such a system is compromised, however, the attacker still faces an egress problem because conventional network paths are unavailable or tightly constrained and information must leave the system through an alternative medium. Covert channels provide such paths by exploiting system resources or observable behavior in ways that violate the intended information flow policy. Prior work has explored covert signaling through network protocols, file systems, timing behavior, and physical-layer effects~\citep{hussain2011high,venvckauskas2013covert,cabuk2009ip,jiang2024physical}. In air-gapped settings, researchers have demonstrated exfiltration through Ethernet cables, keyboard LEDs, speakers, and other peripherals~\citep{guri2021lantenna,guri2019ctrl,choi2023casper,sepetnitsky2014exfiltration}.

USB is a natural interface to examine in this context because it combines power delivery with data transfer and remains available in environments that permit charging or peripheral connectivity. While monitoring of USB data transfers, network activity, and file movement addresses conventional egress paths, it does not directly characterize the observable effects of USB power control. This motivates a closer study of whether USB power control itself can become a covert signaling surface.

In this paper, we present SPECTER, a covert channel that uses host-side USB power control to exfiltrate data from an air-gapped system. We consider a post-compromise setting in which malware has administrative privileges on a host that permits USB charging or peripheral connectivity. The channel assumes that the host and attached device support the required USB control operations and produce distinguishable responses. Obtaining these privileges does not by itself provide an external communication path when conventional network egress is unavailable or tightly constrained.

The transmitter encodes bits through host-side USB power-management or device-state operations without placing the payload in USB data transfers. We evaluate three receiver paths based on the same host-side control mechanism under different attacker assumptions. The first receiver infers bits from charging-state changes on a connected smartphone. In our experiments, an Android application decoded bits from battery-status observations, while the iOS experiment used manual observation of the charging indicator. The second receiver uses line-of-sight observation of externally visible peripheral states, such as LED activity or visible operating transitions. The third receiver observes near-field electromagnetic activity associated with host-side USB control operations using a software-defined radio. These receivers are alternatives rather than simultaneous requirements, and they do not require modified peripheral firmware.

Our experiments on Windows~10 and Ubuntu~22.04 demonstrate the channel on commodity systems, with behavior that varies across the tested host, device, and receiver configurations. The reported signaling rates were 0.2--0.25~bit/s for charging-state reception across Android and iOS devices and up to 1.0~bit/s for selected peripherals on Windows. We additionally report near-field RF reception in a controlled setup with the antenna in direct contact with the USB cable. These low rates motivate focusing on short, high-value secrets rather than bulk data transfer.

These results identify host-side USB power control as an additional exfiltration path and motivate treating USB power-management and device-state events as security-relevant telemetry. The channel can produce host notifications, charging indications, or peripheral interruptions, so its visibility depends on device behavior and the defender's observation capabilities.

Our contributions are as follows.
\begin{itemize}
    \item We introduce SPECTER, a covert channel that uses host-side USB power control to transmit information without encoding the payload in USB data transfers.
    \item We demonstrate three alternative receiver paths based on charging-state observation, peripheral-state observation, and near-field RF observation, and describe the access requirements of each path.
    \item We evaluate the channel on Windows and Ubuntu testbeds and report signaling rates and device-dependent reception outcomes, including configurations in which reception was not obtained.
\end{itemize}

\section{Background and Related Work}
A covert channel exploits system mechanisms to transfer information in violation of established security policies~\citep{lampson1973note,carrara2016air}. In air-gapped environments, where conventional network egress is unavailable, researchers have explored covert communication through thermal, acoustic, electromagnetic, optical, and electrical signals. SPECTER belongs to the power-based category~\citep{zhao2018powermitter,guri2019powerhammer} and uses host-side USB power control to generate observable charging-state, peripheral-state, and RF responses.

\begin{table*}[pos=t]
\centering
\caption{Comparison of representative physical covert channels. Load modulation denotes deliberate variation of computational workload to generate the signal. USB data encoding denotes the use of USB data-transfer payloads to encode the covert signal. Routine USB control activity is not included in this classification. Rates are reported under the respective studies' experimental conditions. The SPECTER rate covers charging-state and peripheral-state reception.}
\label{tab:rel}
\begingroup
\small
\setlength{\tabcolsep}{4pt}
\renewcommand{\arraystretch}{1.15}
\begin{tabularx}{\textwidth}{@{}>{\raggedright\arraybackslash}p{0.25\textwidth}>{\raggedright\arraybackslash}p{0.17\textwidth}>{\raggedright\arraybackslash}p{0.13\textwidth}cc>{\raggedright\arraybackslash}X@{}}
\toprule
\textbf{Paper} & \textbf{Signal source} & \textbf{Reported rate} & \shortstack{\textbf{Load}\\\textbf{mod.}} & \shortstack{\textbf{USB data}\\\textbf{encoding}} & \textbf{Receiver} \\
\midrule
BitWhisper \newline \citep{guri2015bitwhisper} & CPU heat & 1--8 bit/hr & \cmark & \xmark & Thermal sensor \\ \addlinespace[2pt]
MAGNETO \newline \citep{guri2021magneto} & CPU magnetic field & Up to 5 bit/s & \cmark & \xmark & Smartphone magnetometer \\ \addlinespace[2pt]
MOSQUITO \newline \citep{guri2018mosquito} & Speakers & 10--166 bit/s & \xmark & \xmark & Repurposed speakers or headphones \\ \addlinespace[2pt]
CASPER \newline \citep{choi2023casper} & Internal speaker & Up to 20 bit/s & \xmark & \xmark & Microphone \\ \addlinespace[2pt]
USBee \newline \citep{guri2016usbee} & USB data bus RF & 160--640 bit/s & \xmark & \cmark & SDR \\ \addlinespace[2pt]
\citet{spolaor2017no} & Phone power consumption & 1.25 bit/s & \cmark & \xmark & Charging-side power monitor \\ \addlinespace[2pt]
CovertPower \newline \citep{spolaor2025covertpower} & Phone power consumption & Up to 10 bit/s & \cmark & \xmark & Charging-side power monitor \\ \addlinespace[2pt]
PowerHammer \newline \citep{guri2019powerhammer} & Mains current & 10 / 1000 bit/s & \cmark & \xmark & Power-line current sensor \\
\midrule
\textbf{SPECTER} & \textbf{USB power control} & \textbf{0.2--1.0 bit/s} & \xmark & \xmark & \textbf{Phone, visual observer, or SDR} \\
\bottomrule
\end{tabularx}
\endgroup
\end{table*}

\subsection{Air-gapped Covert Channels}
For comparison, we distinguish between techniques that generate signals through deliberate computational load modulation and those that directly control peripheral functions. In the former group, the transmitter varies the workload of computing resources, such as the CPU, GPU, or memory subsystem, to produce physical signals. For instance, Guri et al. introduced BitWhisper, which varies CPU activity to generate thermal signals that can be sensed by nearby systems equipped with thermal sensors~\citep{guri2015bitwhisper}. Similarly, MAGNETO modulates CPU workload to generate low-frequency magnetic fields that can be captured by a magnetometer embedded in a nearby smartphone~\citep{guri2021magneto}. PowerHammer also modulates CPU workload, producing current variations that can be measured on electrical wiring~\citep{guri2019powerhammer}.

Other approaches generate signals by directly controlling peripheral functions rather than modulating computational workload. Guri et al.'s MOSQUITO enables near-ultrasonic communication by reconfiguring connected passive speakers or headphones as input devices, allowing reception without a dedicated microphone~\citep{guri2018mosquito}. Choi et al. proposed CASPER, which uses an internal motherboard speaker to generate inaudible acoustic signals for covert data transfer~\citep{choi2023casper}. These techniques demonstrate that peripheral functions can provide signaling mechanisms without deliberate computational load modulation. Their detectability depends on the resulting physical signals and the available monitoring mechanisms. Table~\ref{tab:rel} summarizes representative channels and their signal sources, reported rates, and receiver requirements.

\subsection{Covert Channels Using USB}
USB provides both a data path and a power path, allowing distinct forms of covert signaling. One representative approach is USBee by Guri et al., which uses the USB data bus as an RF transmitter without requiring hardware modification~\citep{guri2016usbee}. USBee generates electromagnetic emissions through intentional data transmission patterns over the USB interface, and a nearby RF receiver captures these emissions. Its signal generation therefore depends on USB data transfers, whereas SPECTER encodes information through host-side USB power control.

A separate line of work exploits USB power rather than USB data. Spolaor et al. demonstrated data exfiltration from a smartphone to a compromised charging station by inducing CPU activity that produces observable power consumption patterns~\citep{spolaor2017no}. CovertPower further investigated this approach by modulating computational and memory workloads on Android devices and recovering the resulting patterns through a power monitor at the charging side~\citep{spolaor2025covertpower}. In the demonstrated exfiltration paths of both studies, the mobile device is the transmitter and the charging-side power monitor is the receiver. The target data therefore resides on the phone. Spolaor et al. also discuss reverse signaling from the charging source to the phone for a handshake, but leave its implementation to future work~\citep{spolaor2017no}.

SPECTER differs from these approaches in three respects. First, the transmitter is the air-gapped host, and the channel carries data resident on that host. Second, the host generates the signal through USB power-management or device-state operations rather than by inducing computational workload on a mobile device. Third, SPECTER provides three alternative receiver paths based on charging-state observation, peripheral-state observation, and near-field RF observation. Together, these properties establish a new covert channel for exfiltrating host-resident data through USB power control without encoding the payload in USB data transfers.

\subsection{Universal Serial Bus}
USB combines data transfer and power delivery. USB 2.0 signaling uses the D+ and D- lines, while SuperSpeed operation uses additional differential pairs. VBUS and GND provide the power path, with a nominal 5\,V supply in the configurations considered here~\citep{he2015usb}. SPECTER conveys information through observable responses to host-side USB power control without placing the payload in USB data transfers.

A Type-A downstream port supplies power through VBUS. Under standard USB data-port power rules, configured USB 2.0 and USB 3.x devices can draw up to 500\,mA and 900\,mA, respectively, with additional charging capabilities defined separately. Type-C introduces CC pins for attachment detection, role identification, and current advertisement, and can support higher power through USB Power Delivery. Our implementation uses the host-side USB control operations described in Section~\ref{sec:trans} and does not rely on USB Power Delivery negotiation. The effects of these operations depend on the controller, hub, driver, and connected device rather than the connector type alone. In our setup trials, we did not observe qualitative differences between the tested Type-A and Type-C configurations.

The symbol intervals reported in this work correspond to the tested host and receiver configurations. The channel does not encode payload data on the D+ and D- lines or the SuperSpeed pairs. It also does not use Type-C CC or SBU signaling to encode the payload.

\section{Overview}

\begin{figure}[pos=t]
\centering
\includegraphics[width=\columnwidth]{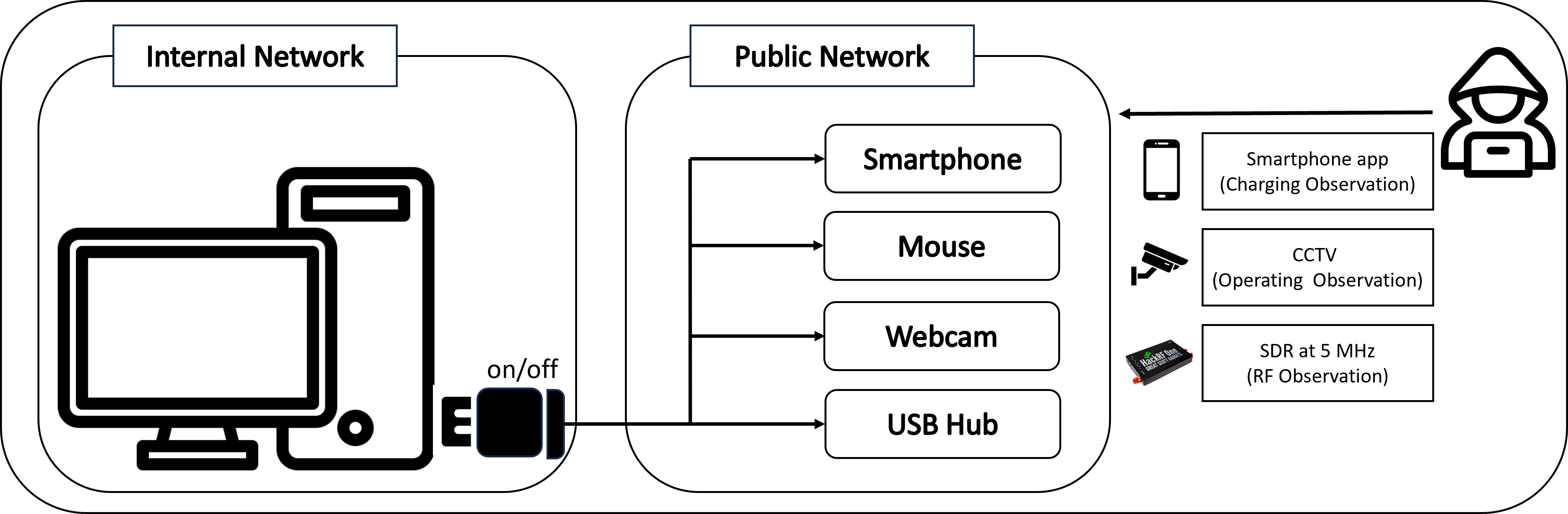}
\caption{Overview of SPECTER. Malware with administrative privileges applies host-side USB power-management or device-state operations. The resulting signals are observed through one of three alternative receiver paths based on smartphone charging states, visible peripheral behavior, or near-field RF activity.}
\label{fig:overview}
\end{figure}

\subsection{Motivation}
Network segregation is used in environments with stringent security requirements to reduce exposure to external threats~\citep{howard1997analysis,patterson2023learning}. Internal systems are physically or logically separated from external networks, restricting the communication paths available to an attacker. Once a host is compromised, however, the effectiveness of this separation also depends on whether other interfaces can carry information across the security boundary.

USB power control provides one such interface in environments that permit USB charging or peripheral connectivity. A compromised host can apply USB power-management or device-state operations that produce observable changes in charging states, peripheral behavior, or electromagnetic activity in responsive configurations. SPECTER uses these effects to establish a covert channel that carries information without encoding the payload in network traffic or USB data transfers. The channel targets small, high-value secrets that can be transmitted at the rates reported in our experiments.

\subsection{Threat Model}
\label{sec:threat-model}
We consider a post-compromise adversary that has obtained administrative privileges on a host in an air-gapped or network-segregated environment where USB charging or peripheral connectivity remains permitted. Administrative privileges refer to root privileges on Linux or administrator privileges on Windows. Malware may reach such a host through infected removable media, supply-chain compromise, or insider-assisted actions~\citep{nissim2017usb,masood2021taxonomy,langner2011stuxnet}. Initial compromise and privilege escalation are outside the scope of this paper. We focus on the subsequent exfiltration stage, when the adversary controls the host but conventional network egress is unavailable or tightly constrained.

\subsubsection{Adversary Goal}
The adversary aims to exfiltrate a small amount of high-value information stored on the compromised host through host-side USB power control. Target information includes short secrets such as cryptographic keys, credentials, or authentication tokens.

\subsubsection{Adversary Capabilities}
The adversary can invoke USB power-management or device-state controls on the target host. The resulting observable responses depend on the controller, hub, driver, and attached device. For the charging-state and peripheral-state paths, the sender uses state-based encoding in which the active control state represents bit~1 and the inactive control state represents bit~0. For the RF path, the sender uses event-based encoding in which one change in the requested USB state within a symbol interval represents bit~1 and the absence of a change represents bit~0. The channel requires no modification to peripheral hardware or firmware and does not encode the payload in USB data transfers.

\subsubsection{Defender Capabilities}
The defender enforces network segregation and policies governing USB connectivity. We consider monitoring of network activity, file transfers, and USB data traffic, while continuous measurement of VBUS waveforms or near-field emissions is outside the assumed routine monitoring. Host logs, charging notifications, and peripheral state changes remain observable. The adversary is not assumed to suppress these observations, and a defender that monitors USB power-management or device-state events may detect the signaling.

\subsubsection{Receiver Models}
We consider three alternative receiver models, illustrated in Fig.~\ref{fig:overview}. Each model provides a separate observation path for the effects of host-side USB power control.

\paragraph{Charging-state receiver.}
A smartphone is connected to the compromised host for charging, and the receiver infers the transmitted bits from its charging state over time. In our Android implementation, an attacker-controlled user-level application decodes bits from battery-status observations without requiring root privileges or modifications to the operating system or firmware. In the iOS evaluation, charging-state changes were observed manually through the charging indicator. The application-based variant therefore requires attacker-controlled software on the receiving phone, whereas the manual variant requires visual access to its charging indicator.

\paragraph{Peripheral-state receiver.}
A benign USB peripheral remains attached to the host, and the receiver infers the transmitted bits from externally visible state changes, such as LED activity or changes in device operation. This path requires a line of sight to the device, which may be provided by a nearby observer or a camera. It does not require attacker-controlled software or modified firmware on the peripheral.

\paragraph{RF receiver.}
An antenna and software-defined radio observe near-field electromagnetic activity associated with host-side USB control operations. This path does not require attacker-controlled software on the attached USB device, but it requires physical proximity to the cable or port area. In the evaluated configuration, the loop antenna was placed in direct contact with the USB cable.

\subsubsection{Attack Scenario}
After compromising the host, the adversary encodes host-resident data through host-side USB power control and observes the resulting signal through one of the receiver paths above. The channel behavior depends on the host configuration, connected device, and observation path. The control operations may also produce host USB notifications, visible charging indications, LED changes, or interruptions in peripheral operation. Where these effects are noticeable, the adversary requires a transmission opportunity during which they do not trigger intervention, such as a period of unattended operation. The threat model therefore distinguishes the ability to transmit information without encoding the payload in USB data transfers from the ability to conceal the associated control operations and device responses.

\section{Host-Side USB Power Control}
\label{sec:trans}

All receiver paths in SPECTER share a host-side USB control mechanism. After compromising the host and obtaining root or administrator privileges, the transmitter applies a sequence of USB power-management or device-state operations. In responsive configurations, these operations produce changes that can be observed through smartphone charging states, peripheral behavior, or near-field RF activity. The platform-specific commands and device paths are provided in the appendix and the artifact repository. The payload is conveyed through these observable effects and is not encoded in USB data transfers.

On Ubuntu~22.04, the transmitter writes to USB power-management attributes under \texttt{sysfs} with root privileges. The released implementation changes \texttt{power/control} and the associated autosuspend settings. Setting \texttt{power/control} to \texttt{on} resumes the device and prevents autosuspend, whereas setting it to \texttt{auto} permits the kernel to suspend an idle device.\footnote{Linux USB power-management documentation is available at \url{https://docs.kernel.org/driver-api/usb/power-management.html}.} These settings control runtime power management and do not by themselves establish that VBUS has been physically disconnected. The resulting behavior depends on the host controller, hub, driver, and connected device. Our tested smartphone configurations required different control sequences, and some Android configurations required a non-default USB mode before the intended charging-state changes could be observed.

On Windows~10, the transmitter issues privileged device enable and disable commands through DevCon, using the identifier of the target USB device.\footnote{We use DevCon, available from Microsoft at \url{https://learn.microsoft.com/en-us/windows-hardware/drivers/devtest/devcon}.} The released implementation invokes the enable command twice because a single invocation did not consistently restore the observable device state in our setup. DevCon controls the operating-system device state, and the resulting charging or peripheral response remains configuration dependent. Device disablement alone is not treated as evidence that VBUS has been physically removed.

The transmitter processes the supplied bit sequence serially and waits after completing the control operations for each bit. The released script uses a five-second waiting period in its Linux smartphone branches and a one-second waiting period in its Windows branches. These are software waiting parameters. The elapsed interval between successive symbols also includes command execution and scheduling delays. Normal USB control activity may accompany these operations, but the transmitter does not place the exfiltrated payload in USB data transfers.

\section{Methodology}
\label{sec:methodology}
SPECTER uses the host-side controls described in Section~\ref{sec:trans} with three alternative observation paths. Charging-state and peripheral-state reception use state-based encoding, while RF reception uses event-based encoding. The encoding rules define how an input bit sequence is mapped to observable states or events. The receiver implementation and its observation requirements differ across the three paths.

\subsection{Covert Channel Using Charging-State Observation}
This path applies when a smartphone is connected to the host for charging and exposes distinguishable charging states in response to host-side control. The sender maps bit~1 to the active control state and bit~0 to the inactive control state, with a waiting period after each operation. Reception uses the corresponding charging-state observations. Consecutive identical bits are represented by maintaining the same signaling state across successive symbol periods.

In the released Android implementation, an attacker-controlled user-level application repeatedly reads the battery status reported by the mobile operating system and decodes successive observation windows into bits. A window is decoded as bit~1 if all sampled states indicate charging or a full battery, and as bit~0 otherwise. The user specifies the number of bits to receive, and the application displays the decoded sequence. The application does not require root privileges or modifications to the phone's operating system or firmware. In the iOS evaluation, charging states were observed manually through the on-screen charging indicator. The length of the observation window is a configuration parameter of the application. The released build takes six samples one second apart, and the experiments used window lengths matched to the transmitter symbol periods reported in Table~\ref{tab:result}.

\subsection{Covert Channel Using Peripheral-State Observation}
This path applies to benign, unmodified USB peripherals whose externally visible behavior changes in response to the transmitter's control operations. Examples include LED activity and changes in device operation. The sender uses the same state-based encoding as the charging-state path, and an external observer interprets the corresponding visible state sequence.

The mapping between a control operation and an observable state is device dependent. Some peripherals expose clear changes, while others remain visibly unchanged or recover too slowly to provide distinguishable symbols at the tested intervals. Reception therefore requires a suitable peripheral and a line of sight to the relevant indicator or behavior. The evaluation uses the visible responses of the tested devices rather than assuming that every device-disable operation removes VBUS power.

\subsection{Covert Channel Using Near-Field RF Observation}

\begin{figure}[pos=!t]
\centering
\includegraphics[scale=0.35]{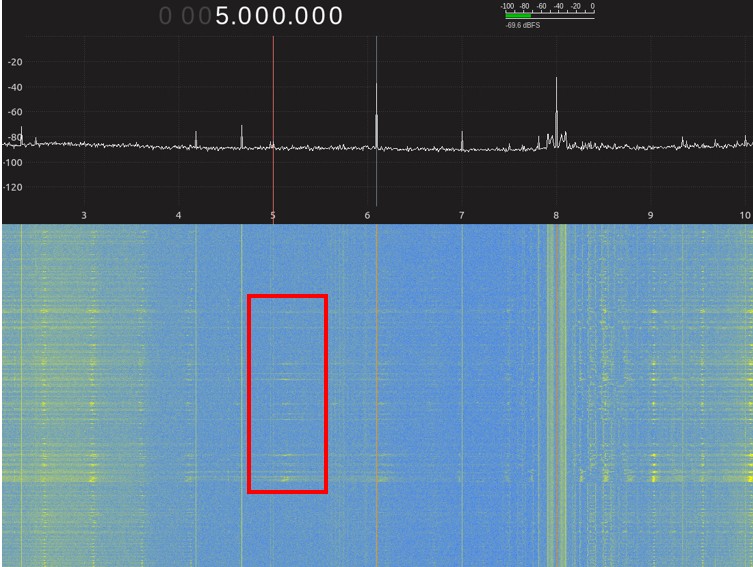}
\caption{GQRX spectrum and waterfall display during host-side USB control. The red box highlights the observation region near 5~MHz.}
\label{fig:gqrx}
\end{figure}

Host-side USB control also produces changes in electromagnetic activity that can be observed near the USB cable. We inspected the GQRX spectrum and waterfall display and selected a region near 5~MHz in which switching-related variations were visually distinguishable in our setup, as illustrated in Fig.~\ref{fig:gqrx}. This frequency is a configuration-specific observation point. The evaluated receiver uses a HackRF One and an LF-R 400 loop antenna placed in direct contact with the USB cable.

The released RF analysis script constructs a complex sequence from interleaved I/Q samples and applies a short-time Fourier transform with a sampling rate of 2.6~MS/s and a segment length of 8192 samples. It extracts the magnitude of the zero-frequency STFT bin and discards the first and last 20 time frames for visualization. For the retained frames, the plotted quantity is
\begin{equation}
A_k = |Z(0,k)|,\qquad R_k = \frac{A_{\max}-A_k}{A_{\max}},
\end{equation}
where $Z(0,k)$ is the STFT coefficient at the zero-frequency bin in frame $k$ and $A_{\max}$ is the maximum retained magnitude. This processing produces a dimensionless temporal trace for examining variations associated with the transmitter's control sequence.

The RF path uses event-based encoding because the presence of a switching-related burst is the observation of interest. For bit~1, the sender changes the requested USB state once at the beginning of the symbol. For bit~0, it leaves the requested state unchanged. The decoding rule assigns bit~1 to an interval containing the corresponding burst and bit~0 to an interval without it. The Windows RF branch uses a one-second waiting period after each symbol's control actions, so the waiting parameter does not establish an exact one-second interval between physical events.

The released RF script implements trace extraction and visualization. It does not implement automatic frame synchronization, threshold selection, or binary decision output. The event-based decoding rule is therefore distinguished from the functionality provided by that script. The RF observations characterize the contact-probe configuration used in the evaluation.

\subsection{Packet Format}

\begin{figure}[pos=!t]
\centering
\includegraphics[scale=0.35]{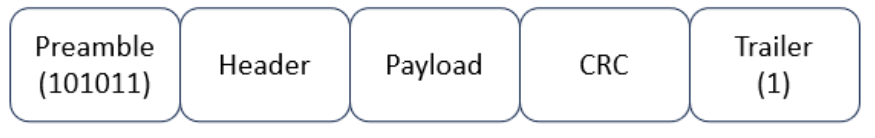}
\caption{Packet structure for framed transmissions using charging-state or peripheral-state observation. The header contains the payload length.}
\label{fig:packet}
\end{figure}

The transmitter accepts a binary sequence prepared before transmission. Depending on the experiment, this input consisted of either a preassembled packet or a test sequence without the complete packet structure. Framing is performed before the sequence is supplied to the transmitter, which applies the selected encoding rule without interpreting individual packet fields.

For framed transmissions using charging-state or peripheral-state observation, the packet consists of a preamble, a header containing the payload length, a payload, a CRC, and a trailer, as shown in Fig.~\ref{fig:packet}. The preamble \texttt{101011} provides a known pattern for identifying the packet start. The header specifies the payload size, and the payload contains the bits to be exfiltrated. No additional compression or encryption layer is applied by the transmitter.

The CRC field is included for error detection~\citep{ramabadran1988tutorial}. The channel provides no acknowledgment path or feedback-based retransmission. In the state-based format, the trailer is set to bit~1 so that the transmitter requests the active USB state after the packet.

The RF path uses the shorter preamble \texttt{101} when framing is applied, followed by event-based payload symbols. The same event-based encoding rule also applies to test sequences supplied without the complete framing structure. Packet structure and symbol encoding are therefore separate parts of the transmission procedure.

\section{Evaluation}
\subsection{Experimental Setup}
We evaluated the channel on Windows~10 and Ubuntu 22.04.4 64-bit, using a desktop for Ubuntu and a laptop for Windows. Both machines were equipped with USB 2.0, USB 3.0 Type-A, and USB Type-C ports. All experiments ran in a controlled laboratory without network connectivity, and each receiver path was evaluated separately.

For charging-state observation, we used a Galaxy S22 running Android 14 and an iPhone 13 Pro running iOS 17.4.1. A custom Android application observed the battery status reported by the operating system and decoded the corresponding bit sequence on the Galaxy S22. For the iPhone, we manually observed the on-screen charging indicator. For peripheral-state observation, we attached one USB peripheral at a time and observed its visible behavior, testing a USB mouse, keyboard, webcam, wireless-headset dongle, and USB hub. For near-field RF observation, we used a HackRF One with an LF-R 400 loop antenna placed in direct contact with the USB cable.

Table~\ref{tab:result} summarizes the reception outcomes and reported signaling rates for the tested host and device configurations. These signaling rates are distinct from net payload goodput after packet framing overhead.

\begin{table}[t]
\centering
\caption{Reception outcomes and reported signaling rates for the tested host and device configurations. N/O indicates that we did not obtain a usable channel in the tested configuration.}
\label{tab:result}
\resizebox{\columnwidth}{!}{%
\begin{tabular}{llcc}
\toprule
\multicolumn{2}{c}{\textbf{Receiver path}} & \textbf{Ubuntu desktop} & \textbf{Windows laptop} \\
\midrule
\multirow{2}{*}{Charging-state} & Android smartphone & 0.2 bit/s & 0.25 bit/s \\
                              & iPhone & 0.2 bit/s & 0.25 bit/s \\
\midrule
\multirow{5}{*}{Peripheral-state} & Mouse & N/O & 1.0 bit/s \\
                                & Keyboard & N/O & N/O \\
                                & Webcam & N/O & 0.25 bit/s \\
                                & Headset dongle & N/O & N/O \\
                                & USB hub & N/O & 0.25 bit/s \\
\bottomrule
\end{tabular}%
}
\end{table}

\subsection{Charging-State Receiver}

We observed usable reception on both Android and iOS in the tested Ubuntu and Windows configurations. The reported signaling rates were 0.2~bit/s on the Ubuntu desktop and 0.25~bit/s on the Windows laptop for both smartphones, as summarized in Table~\ref{tab:result}. Since the operating systems were evaluated on different host machines, the rate difference cannot be attributed to the operating system alone.

On some device and platform combinations, repeated control operations produced visible charging indications and brief screen activity. In the fully charged condition observed in our tests, the operations did not wake the screen or produce a visible charging indication. This observation concerns the screen behavior of the tested configuration and does not establish a general reduction in channel detectability.

\subsection{Peripheral-State Receiver}

On the Windows laptop, we obtained usable channels with a mouse at 1.0~bit/s, a webcam at 0.25~bit/s, and a USB hub configuration at 0.25~bit/s. On the Ubuntu desktop, we did not obtain a usable channel with any of the tested peripherals.

\paragraph{Mouse.}
The mouse provided the clearest visible response on the Windows laptop. When the target device was disabled, it stopped responding and its LED turned off. The resulting state changes supported signaling at the reported rate of 1.0~bit/s.

\paragraph{Webcam.}
The webcam supported signaling at 0.25~bit/s on the Windows laptop. Its status LED was active only while a camera application was using the device, so reception through that indicator required the webcam to be in use. This behavior was observed for the tested external and built-in webcams.

\paragraph{Headset dongle.}
The USB dongle could be disabled on Windows, but audio playback continued. The observed behavior therefore did not provide a distinguishable binary state for reception in the tested configuration.

\paragraph{USB hub.}
The hub's LED remained on during the control operations, while downstream devices stopped functioning. Reception therefore relied on the behavior of the downstream devices rather than the hub's own LED. This configuration supported signaling at the reported rate of 0.25~bit/s on the Windows laptop.

\paragraph{Keyboard.}
The keyboard did not yield repeatable, distinguishable state changes under the tested control operations on either host.

These results indicate that peripheral-state reception depends on the combination of host configuration and observable device behavior. A device-control operation that produces a useful visible signal for one peripheral may not produce one for another.

\subsection{Near-Field RF Receiver}

\begin{figure}[pos=t]
\centering
\includegraphics[width=\linewidth]{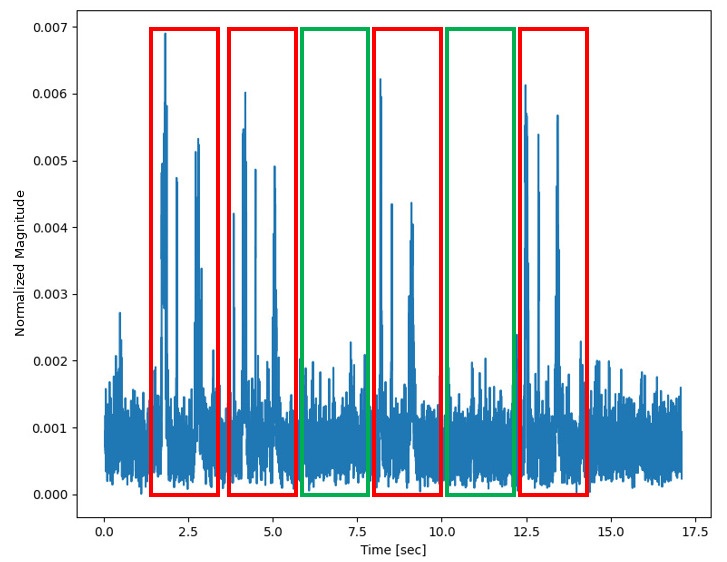}
\caption{RF trace recorded near 5~MHz during transmission of the 6-bit sequence \texttt{110101}. Red and green boxes were manually added to illustrate intervals corresponding to bits~1 and~0, respectively.}
\label{fig:110101}
\end{figure}

\begin{figure}[pos=t]
\centering
\includegraphics[width=\linewidth]{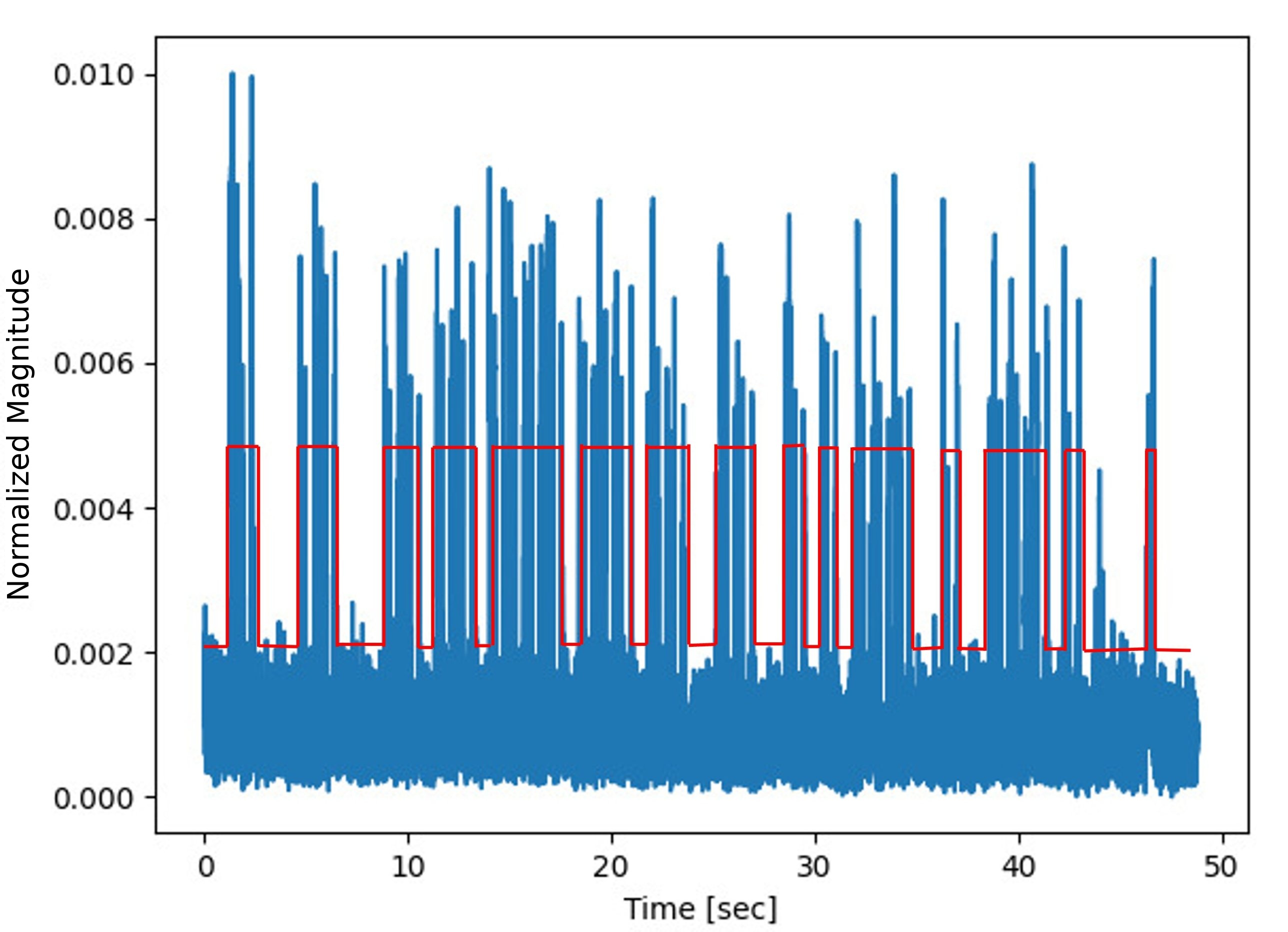}
\caption{RF trace recorded near 5~MHz during transmission of a 48-bit payload. The blue trace shows the processed RF recording, and the red line was added manually as a visual guide.}
\label{fig:covert_receive}
\end{figure}

\begin{figure}[pos=t]
\centering
\includegraphics[width=\linewidth]{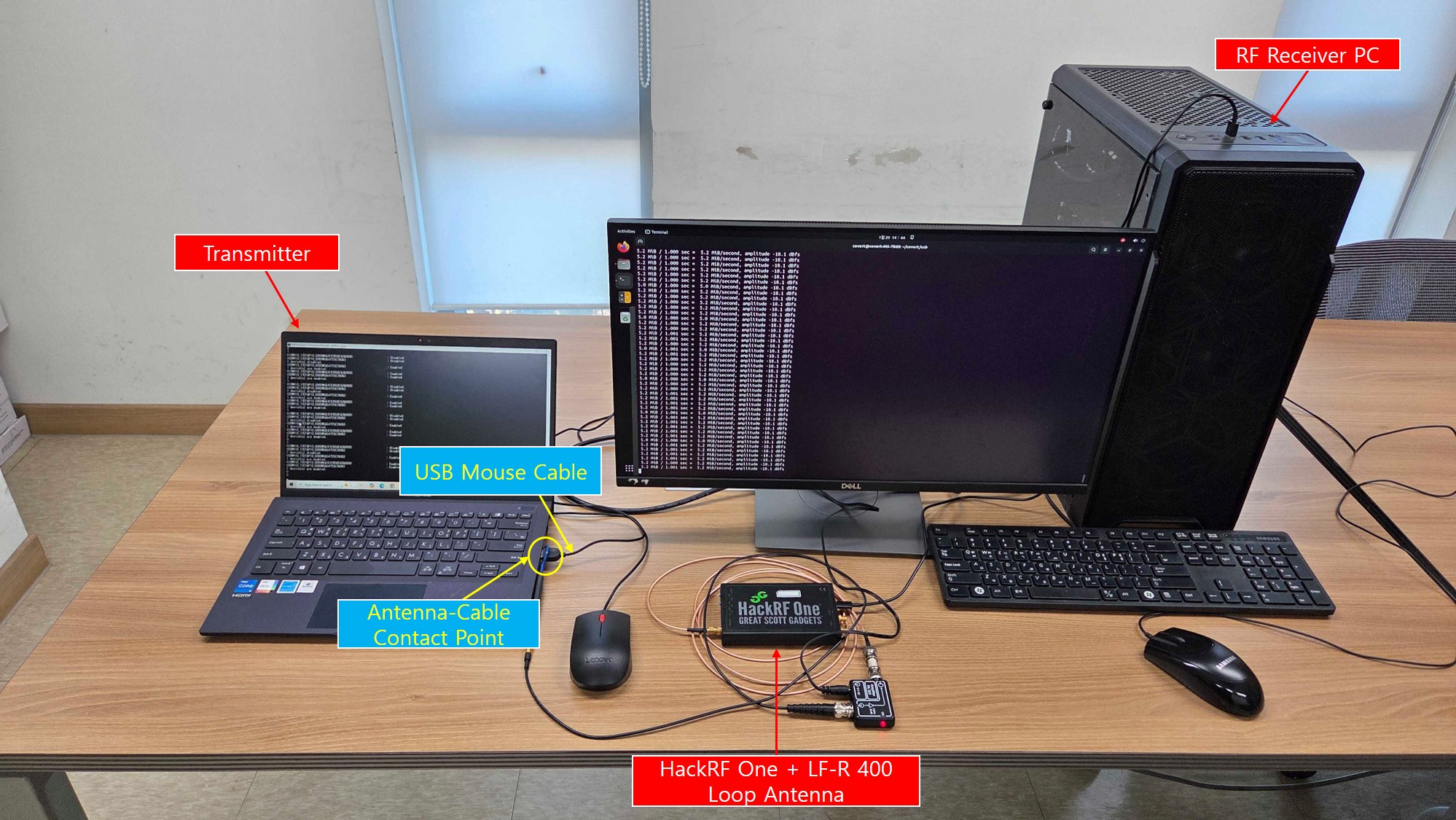}
\caption{Near-field RF receiver setup. The transmitter laptop on the left runs the transmission software on Windows with a USB mouse attached. A HackRF One with an LF-R\,400 loop antenna is placed in direct contact with the USB mouse cable. The desktop on the right records the RF signal on Ubuntu.}
\label{fig:rf-setup}
\end{figure}

We evaluated RF reception using electromagnetic activity observed during host-side USB control. The loop antenna was placed in direct contact with the USB mouse cable, as shown in Fig.~\ref{fig:rf-setup}. The results therefore apply to this contact-probe configuration.

Bits were identified from the presence or absence of switching-related bursts distinguishable from background activity within the corresponding symbol intervals. Figure~\ref{fig:110101} shows reception of the 6-bit sequence \texttt{110101}, and Fig.~\ref{fig:covert_receive} shows reception of a 48-bit payload. The vertical axes in these figures represent the normalized signal magnitude defined in Eq.~(1). In these two examples, the decoded sequences matched the transmitted sequences, including the longer example with brief noise bursts in the recording.

The absence of observed errors is specific to these examples and does not establish a channel-wide error rate. Reception at increasing probe distances, sensitivity to antenna orientation, and robustness under different environmental conditions were not characterized.

\section{Discussion}

SPECTER introduces a covert channel that uses host-side USB power control to transmit information from a compromised host. The three receiver paths demonstrate how charging states, peripheral behavior, and near-field RF activity can convey information without placing the payload in USB data transfers. The security implication is that monitoring USB data traffic alone does not account for all signaling opportunities exposed by the USB subsystem. Host-side control operations and their externally observable effects therefore warrant consideration in USB security policies.

\subsection{Limitations}

SPECTER assumes a post-compromise adversary with root or administrator privileges. Initial compromise, privilege escalation, and persistence are outside the scope of this work. The channel applies to environments in which conventional network egress is unavailable or tightly constrained while USB charging or peripheral connectivity remains permitted.

Each receiver path has distinct operational requirements. Android charging-state reception requires a connected smartphone with an attacker-controlled user-level application, while the evaluated iOS path relies on manual observation of the charging indicator. Peripheral-state reception requires a device that exposes distinguishable behavior and a line of sight to that behavior. The evaluated RF path requires an SDR and a loop antenna in direct contact with the USB cable. Reported signaling rates for the charging-state and peripheral-state paths range from 0.2 to 1.0~bit/s, limiting the amount of information that can be transmitted within a given time.

The experiments were conducted in a controlled laboratory using selected host and device configurations. The same control operations and receiver responses may not generalize to other controllers, drivers, or peripherals. Repeated control operations can also produce host notifications, charging indications, screen activity, or device interruptions. Detectability therefore depends on the observable effects of the selected configuration and the monitoring available to the defender. The evaluation reports signaling rates and individual reception examples but does not provide systematic estimates of bit error rate, packet error rate, or effective payload goodput.

\subsection{Defensive Implications}

The results motivate treating USB power-management and device-state events as security-relevant telemetry. A candidate monitoring approach is to audit repeated enable, disable, suspend, and resume operations involving the same device. Correlating these events with application activity and available charging or peripheral-state observations may help identify patterns that warrant investigation. Such correlation would need to distinguish deliberate signaling from legitimate device management, reconnection, and power-saving behavior.

Policy measures may include restrictions on software-controlled USB state changes, limits on unused ports, device allowlists, and restrictions on personal smartphones near sensitive hosts. Because SPECTER can use benign peripherals, allowing a device to connect does not by itself address the subsequent use of host-side control operations for signaling. The effectiveness of host-based monitoring and enforcement also depends on whether those mechanisms remain trustworthy after privileged compromise.

Prior studies of anomaly detection in IoT networks provide general examples of learning-based monitoring~\citep{salman2022anomaly,inuwa2024comparative}. Applying such approaches to USB event sequences would require appropriate features, device-specific baselines, and evaluation under normal operating workloads. The present experiments do not validate a detector, a rate-limiting policy, or a universal decision threshold. These measures are candidate defenses whose effectiveness and operational costs require separate evaluation.

\subsection{Future Work}

Future work should characterize reliability and usable throughput through repeated transmissions across a broader range of hosts, controllers, and devices. Relevant measurements include bit and packet error rates, synchronization failures, performance with longer payloads, and payload goodput after framing overhead. Alternative encoding and framing schemes could also be examined for their effects on synchronization, throughput, and visible device behavior.

For RF reception, further work should examine probe distance, placement, orientation, and environmental interference. For defense, host-level monitoring approaches should be evaluated against both deliberate signaling and legitimate USB activity in enterprise and industrial workloads. Such evaluation should consider false positives, effects on normal device operation, and the integrity of monitoring under the assumed privileged adversary.

\subsection{Ethical Considerations}

All experiments were conducted on researcher-owned devices in an isolated laboratory. No third-party systems, data, or infrastructure were targeted, and no unauthorized access was performed. The released artifact contains a host-side transmitter, an Android charging-state receiver, and an RF trace-processing script. Device identifiers require explicit configuration before the transmitter is used. These materials support inspection of the implementation and further research on USB power-based signaling.

The study examines the security implications of host-side USB control capabilities and their observable effects. It does not present the channel as a vendor-specific software defect. The purpose of the work is to improve understanding of this signaling mechanism and inform monitoring and policy decisions in environments that rely on network isolation.
\section{Conclusion}
This paper presented SPECTER, a covert channel that uses host-side USB power control to exfiltrate information from a compromised air-gapped or network-segregated host. Assuming malware with root or administrator privileges, SPECTER conveys information through the observable effects of USB control operations without encoding the payload in USB data transfers. We demonstrated information recovery through three alternative receiver paths based on charging-state observation, peripheral-state observation, and near-field RF observation. Experiments on Windows~10 and Ubuntu~22.04 demonstrated the channel on commodity systems. The reported signaling rates were 0.2--0.25~bit/s for charging-state reception and up to 1.0~bit/s for selected peripherals on the Windows host. RF reception was demonstrated with a loop antenna in direct contact with the USB cable. These results show a low-rate signaling capability relevant to the exfiltration of short secrets, with behavior dependent on the host configuration, connected device, and receiver path. SPECTER identifies host-side USB power control as a signaling surface that warrants consideration in the security of isolated systems. The findings motivate monitoring USB power-management and device-state operations alongside conventional network and USB data activity. Future work should characterize reliability, usable throughput, and detectability across broader hardware and software configurations and evaluate monitoring approaches for suspicious USB control sequences.

\section*{Acknowledgements}
This work was supported by Institute of Information \& communications Technology Planning \& Evaluation (IITP) grant funded by the Korea government (MSIT) (No.2020-0-00913, Study on Wireless Covert Channel Risk Verification). This research was also partially supported by the Korea University Sabbatical Year Support Program.

\appendix
\counterwithin{figure}{section}
\counterwithin{table}{section}
\counterwithin{equation}{section}
\section{USB Power Control Procedures}
\subsection{Ubuntu}
The Ubuntu implementation accesses USB runtime power-management attributes through \texttt{sysfs}. A device path such as \texttt{/sys/bus/usb/devices/1-1} identifies the target device according to the USB bus and port topology. Setting \texttt{power/control} to \texttt{on} resumes the device and prevents autosuspend, whereas \texttt{auto} permits the kernel to suspend an idle device. The resulting charging-state response depends on the connected device, its USB configuration, and the host software stack.

\subsubsection{iOS Devices}
For the iOS branch, the released script writes to attributes under the target device's \texttt{power} directory. The active setting uses \texttt{on} for \texttt{control} and \texttt{2000} for \texttt{autosuspend\_delay\_ms}. The inactive setting uses \texttt{auto} and \texttt{0}, respectively. The script also writes to the legacy attributes \texttt{level} and \texttt{autosuspend}, using \texttt{on} and \texttt{2} for the active setting and \texttt{auto} and \texttt{0} for the inactive setting. Availability of these legacy attributes depends on the kernel. The writes require administrative privileges and are issued through \texttt{sudo tee}.

\subsubsection{Android Devices}
\begin{figure}[pos=t]
\centering
\includegraphics[scale=0.3]{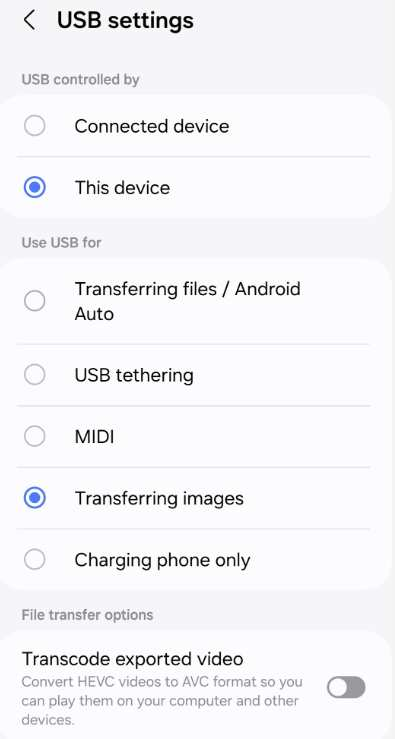}
\caption{USB mode selection on the Samsung Galaxy S22 used in the Ubuntu charging-state evaluation.}
\label{fig:setting}
\end{figure}

The Samsung branch of the released script issues the same attribute writes through both \nolinkurl{/sys/bus/usb/devices/<device>/power/} and \nolinkurl{/sys/bus/usb/drivers/usb/<device>/power/}. These paths can resolve to the same underlying device, so the repeated writes do not imply eight independent power-control attributes. The use of both paths describes the released implementation rather than a general requirement for Android devices. The tested Galaxy S22 configuration also required a non-default USB mode to obtain the charging-state response, as illustrated in Fig.~\ref{fig:setting}.

\subsection{Windows}
The Windows implementation uses DevCon to change the operating-system state of the target USB device. Devices can be listed with \texttt{devcon findall "*USB*"}, and their hardware identifiers can be inspected with \texttt{devcon hwids "*USB*"}. The following commands disable and enable devices matching the specified hardware identifier.

\begin{verbatim}
devcon disable "<HWID>"
devcon enable "<HWID>"
\end{verbatim}

Here, \texttt{<HWID>} is replaced with the target hardware identifier. If a device instance identifier is used instead, it must be prefixed with \texttt{@}. The control commands require administrative privileges. The released transmitter invokes the enable command twice because a single invocation did not consistently restore the observable device state in the tested configuration.

\subsection{Summary of USB Control Operations}
Table~\ref{tab:usb-compare} summarizes the control mechanisms used by the released implementation. Both implementations require administrative privileges. Their observable effects depend on the host controller, driver, and connected device. Successful execution of a power-management or device-state command alone does not establish that VBUS has been physically disconnected.

\begin{table}[t]
\centering
\caption{USB control mechanisms used in the released implementation. The Ubuntu entries describe script branches and access paths rather than universal requirements for each mobile operating system.}
\label{tab:usb-compare}
\resizebox{\columnwidth}{!}{%
\begin{tabular}{ll}
\toprule
\textbf{Platform and branch} & \textbf{Control mechanism} \\
\midrule
Ubuntu, iOS branch & Power-management attribute writes under \\
                  & \quad \texttt{/sys/bus/usb/devices/} \\
\midrule
Ubuntu, Samsung branch & Power-management attribute writes through \\
                       & \quad \texttt{/sys/bus/usb/devices/} \\
                       & \quad \texttt{/sys/bus/usb/drivers/usb/} \\
\midrule
Windows & DevCon device-state commands \\
        & \quad \texttt{devcon disable "<HWID>"} \\
        & \quad \texttt{devcon enable "<HWID>"} \\
\bottomrule
\end{tabular}%
}
\end{table}

\section{Open Science}
All artifacts are available at \url{https://github.com/hjchoi17/SPECTER}, including the transmitter (\texttt{final\_send.py}), the RF receiver (\texttt{RF\_receive.py}), and the Android application (\texttt{charging\_receive.apk}). The README describes setup steps for all three receiver configurations. The transmitter requires explicit configuration of device identifiers to reduce the risk of accidental execution.

\bibliographystyle{cas-model2-names}
\bibliography{bib}

\end{document}